\documentclass{article}
\usepackage{spconf,amsmath,amssymb}
\usepackage{graphicx}
\usepackage{algorithm, algorithmic}
\usepackage{textcomp}
\usepackage{booktabs}
\usepackage{subcaption}
\usepackage{siunitx}
\usepackage{float}

\usepackage{hyperref}
\usepackage{xcolor}

\newcommand{\transpose}[1]{\ensuremath{{#1}^{\textsc{t}}}}
\newcommand{\trace}[1]{\ensuremath{\text{Tr}\left(#1\right)}}

\title{Eccentricity Confound in EEG-based Visual Attention Decoding from Gaze-Fixated Neural Tracking of Motion in Natural Videos}
\name{\begin{tabular}{c}
{Yuanyuan Yao}$^{1,*}$, {Celina Salamanca González}$^1$, {Simon Geirnaert}$^{1,2}$, {Céline R. Gillebert}$^{3,5}$, \\
{Tinne Tuytelaars}$^4$, {Alexander Bertrand}$^1$
\end{tabular}\thanks{This research is funded by the Research Foundation - Flanders (FWO) project No G081722N and G026026N, junior postdoctoral fellowship fundamental research of the FWO (for S. Geirnaert, No. 1242524N), Internal Funds KU Leuven (projects IDN/23/006, C14/25/108, C3/25/107, and METH/24/003), and the Flemish Government (AI Research Program). \\Y. Yao, S. Geirnaert, T. Tuytelaars and A. Bertrand are also affiliated with Leuven.AI - KU Leuven institute for AI, Belgium.}\thanks{$^*$Corresponding author: Yuanyuan Yao (yuanyuan.yao@kuleuven.be).}}
\address{$^1$KU Leuven, Department of Electrical Engineering, \\STADIUS Center for Dynamical Systems, Signal Processing and Data Analytics, Belgium\\
$^2$KU Leuven, Department of Neurosciences, Research Group ExpORL, Belgium\\
$^3$KU Leuven, Brain and Cognition, Belgium\\
$^4$KU Leuven, Department of Electrical Engineering, Processing Speech and Images (PSI), Belgium\\
$^5$Leuven Brain Institute (LBI), Belgium\\
}

\begin{document}
%\ninept
%
\maketitle
\begin{abstract}
\textit{Objective.} Decoding visual attention from brain signals during naturalistic video viewing has emerged as a new direction in brain-computer interface research. Current methods assume that stronger coupling between object motion and neural activity indicates higher attention, but this can be confounded by eye movement artifacts and stimulus properties. This study investigates how visual eccentricity—the distance between a visual object and the fixation point—affects neural responses when eye movement artifacts are controlled.
\textit{Approach.} EEG signals were recorded across three tasks that manipulated object eccentricity and attention conditions while participants maintained gaze fixation. Correlation analysis and match-mismatch decoding were performed to quantify the neural tracking of object motion.
\textit{Main results.} The analysis supports three conclusions: (1) neural tracking of object motion in natural videos works under gaze fixation; (2) the strength of this tracking under gaze fixation is modulated by attention; and (3) there exists a significant eccentricity confound in the EEG responses, with poorer neural tracking of motion at larger eccentricities.
\textit{Significance.} These results indicate that findings from previous free-viewing studies also reflect genuine neural processing rather than mere oculomotor artifacts. However, the identified eccentricity effect highlights a major limitation for current decoding approaches that assume coupling strength reflects attention levels alone.
\end{abstract}

\section{Introduction}
\label{sec:intro}

\par Neural activity is modulated by attention: neurons processing task-relevant information exhibit increased firing rates and reduced variability, while those processing irrelevant information are suppressed \cite{thiele2018neuromodulation}. This enhancement is likewise reflected in non-invasive, population-level electrophysiology, where the sensory-evoked response to an attended stimulus is amplified \cite{hillyard1998event}. While such effects are typically quantified as the amplitude of a time-locked evoked response, the same enhancement can be captured by modeling the ongoing relationship between the stimulus and the neural response, where it appears as stronger temporal coupling between the two \cite{lalor2007isolating, lalor2009resolving}. This coupling has been observed not only with simple synthetic stimuli but also with naturalistic speech, images, and videos \cite{mesgarani2012selective, baldauf2014neural, dmochowski2018extracting, yao2024identifying}.

\par However, attention is not the only factor that influences stimulus-brain coupling strength. When using naturalistic stimuli, various confounding factors can modulate this coupling. For example, Li et al. \cite{li2025temporal} reconstructed speech envelopes from electroencephalography (EEG) signals and used the correlation between reconstructed and actual envelopes as a metric for auditory attention. While this score was higher when participants were attentive and lower during distraction, it also correlated with the time-varying standard deviation of the speech envelope, suggesting that the score was also sensitive to stimulus dynamics rather than attention alone. When this confound was not addressed, decoding performance deteriorated. This exemplifies that stimulus-brain coupling is not a pure measure of top-down goal-directed attention, but is also dependent on bottom-up sensory stimulus properties \cite{corbetta2002control,itti2001computational}. Moreover, such bottom-up influences may not cancel out in complex naturalistic conditions and could systematically bias decoding results in real-world applications.

\par These concerns extend naturally to newly emerging EEG-based visual attention decoding paradigms with potential applications in education, neuromarketing, and clinical diagnostics \cite{dmochowski2018extracting, yao2024identifying, yao2025eeg}. We hypothesize that the spatial location of visual objects relative to gaze position (i.e., eccentricity, calculated from the center of the object's bounding box) could be a potential confound: a higher stimulus-brain coupling may not necessarily indicate higher attention, but could also result from the object being located at a more central location in the visual field. This is motivated by the well-established nonuniformity of visual processing, with maximal visual acuity and pattern recognition capabilities at the fovea and a rapid decline toward the periphery \cite{curcio1990human, strasburger2011peripheral}. A direct link between eccentricity and neural responses has been demonstrated previously. For instance, in a functional magnetic resonance imaging study by Wang et al. \cite{wang2016neural}, participants were presented with images of different categories (faces, houses, etc.) at different eccentricity positions. They reported that the neural activation was most intense when stimuli were in the central visual field, and the magnitude of responses declined as eccentricity increased. 

\par Beyond these stimulus-dependent effects, eye movements present another persistent confound in free-viewing paradigms. Under natural conditions, attention is typically accompanied by overt behavior such as directing gaze toward the attended object, which is known as overt attention. This is particularly relevant for paradigms relying on the motion information in naturalistic videos, as eye movements (saccades and smooth pursuit) are naturally evoked by the movements in the videos, and thus may strongly correlate with the video motion features. These eye movements can influence stimulus-EEG correlations through several mechanisms: (1) (residual) electrooculogram (EOG) artifacts in the EEG signals, (2) neural motor activity associated with the planning and execution of eye movements, and (3) neural responses to gaze-induced visual motion, akin to a moving-camera effect where the eyes serve as the camera. This ambiguity undermines findings from prior studies that reported significant neural tracking of video motion under free-viewing conditions \cite{dmochowski2018extracting, yao2024identifying}, since it remains unclear to what extent these effects reflect genuine neural processing of motion information rather than oculomotor contributions. In \cite{yao2025eeg}, eye gaze data were explicitly shown to be correlated with video motion features, and using gaze data alone could achieve comparable attention decoding performance to EEG-based decoders, which again raises concerns about the neural origin of the reported effects. Similarly, Ki et al. \cite{ki2021neural} showed that the neural responses are spatially selective and are enhanced for task-related locations, but since their paradigm relied on overt attention, doubts remain as to whether the reported enhancement was simply an artifact of eye gaze bias toward task-relevant locations, rather than a genuine neural correlate of covert attentional selection.

\par Since disentangling genuine neural correlates of attention from oculomotor confounds is particularly challenging under free-viewing conditions, we follow a common strategy in cognitive neuroscience: employing a covert attention paradigm. Participants maintain fixation while attending to stimuli in the visual periphery, thereby eliminating eye-movement confounds at the cost of reduced ecological validity. This approach is well established: sustained covert attention to peripheral locations reliably modulates early visual responses, and can even be split across multiple non-contiguous peripheral targets \cite{muller2003sustained}. This controlled approach allows us to rigorously test two key questions: (1) whether neural tracking of video motion persists in the absence of eye movement contributions, and (2) whether eccentricity of visual stimulation systematically influences this tracking.

\par The remainder of this paper is organized as follows. Section~\ref{sec:protocol} describes the experimental protocol, including visual stimuli, tasks, and data acquisition. Section~\ref{sec:method} details preprocessing and correlation analysis methods. Section~\ref{sec:results} presents results on the effects of attention and eccentricity on single-subject decoding performance and inter-subject correlation (ISC). Section~\ref{sec:discussion} discusses the implications of our findings. Finally, Section~\ref{sec:conclusion} concludes the paper.

\section{Experimental Protocol}
\label{sec:protocol}

\par $14$ healthy young adults (6 men, 8 women), primarily university students, participated in the study, each providing written informed consent. All participants had normal or corrected to normal vision and confirmed they had no history of neurological diseases. Three visual attention tasks were designed to investigate the effects of attention and eccentricity confounds. The study was conducted in accordance with the ethical principles of the Declaration of Helsinki, and was approved by the KU Leuven Social and Societal Ethics Committee (Application No. G-2022-4765-R3).

\subsection{Visual content}

\par The stimuli consisted of seven single-shot natural videos taken from the video set used in \cite{yao2024identifying, yao2025eeg}. Each video featured a person performing a stage act and was truncated to \SI{3}{\minute} duration with a frame rate of \SI{30}{\hertz}. To better control the performer's eccentricity relative to the gaze fixation point, we centered the performer in each video by (1) detecting the performer's bounding box in each frame using a pre-trained Mask R-CNN model \cite{he2017mask}\footnote{The Mask R-CNN model used an Inception v2 backbone with pre-trained weights from the COCO (Common Objects in Context) dataset. The model exhibited an exceptionally low bounding-box detection failure rate on our experimental stimuli (ranging from $0\%$ to $1.46\%$ across videos, mean = $0.24\%$). Missing bounding boxes in these frames were resolved via linear interpolation.}, (2) defining the bounding box size as the maximum width and height across all frames, (3) smoothing the bounding box center positions to reduce abrupt movements, and (4) cropping the video frames based on the smoothed centers and fixed bounding box size. The resulting cropped videos varied in size and were resized, with aspect ratios preserved, to a height of 540 pixels for display on a $1920 \times 1080$ pixel canvas with a black background. Given the display geometry (Section~\ref{sec:data_acquisition}), 540 pixels subtended approximately $6^\circ$ of visual angle in height. While the stimulus width varied across individual videos to preserve the original aspect ratios of the cropped bounding boxes, it remained close to the height, with its horizontal visual angle ranging from $6^\circ$ to $7^\circ$.

\par Three tasks were designed (see Section~\ref{sec:tasks}), and the same seven videos were used in all three tasks. In Tasks 1 and 2, each cropped video gradually shifted from the screen center toward either the left or right side at a constant speed of $0.1$ pixels per frame. In contrast, the videos in Task 3 remained centered (see Figure \ref{fig:tasks}). Because spatial direction was not an experimental factor of interest, the direction of the shift in Tasks 1 and 2 was treated as a nuisance variable and randomly assigned to prevent systematic directional bias. At this constant speed, and with a frame rate of 30 frames per second, each video's center translated a maximum of 540 pixels from the screen center over the 3-minute duration. At maximum displacement, the video occupied the visual field from roughly $3^\circ$ to $9^\circ$ eccentricity (inner edge to outer edge, respectively).

\par Throughout all tasks and videos, a fixation cross ($60 \times 60$ pixels) was presented at the screen center. Its luminance decreased and returned to normal at irregular intervals. Additionally, a red circle (radius: $50$ pixels) appeared and faded out at random times and at random locations within each cropped video. An example frame with a circle appearing is shown in Figure \ref{fig:circle_cross}. These two visual events never occurred simultaneously, and each of these fade-out events lasted 2 seconds to avoid inducing strong time-locked event-related potentials. To keep participants engaged, they were instructed to respond to these two different types of events in different tasks (see Section~\ref{sec:tasks}). The timing of these event onsets relative to each video was identical for all participants. During the eccentric videos in Tasks 1 and 2, consecutive crosses appeared on average every 19.9 s (range 10.1 to 42.4 s) and consecutive circles every 24.1 s (range 10.0 to 59.4 s). Events were sparser during the central videos in Task 3, with consecutive crosses spaced an average of 44.0 s apart (range 10.6 to 89.1 s) and consecutive circles spaced 39.3 s apart (range 10.3 to 100.8 s). For both conditions, these intervals were highly variable, ensuring that the appearance of crosses and circles remained temporally unpredictable rather than rhythmic. Because these event sequences were fixed per condition, every participant experienced the exact same total number of events throughout the experiment: 64 cross fade-outs and 43 circle events in Tasks 1 and 2, and 22 cross fade-outs and 21 circle events in Task 3.
\begin{figure}[!t]
  \centering
  \includegraphics[width=0.45\textwidth]{circle_cross_blurred.png}
  \caption{Illustration of a frame showing the central fixation cross and a red circle appearing at a random location in the video. The face is blurred for identification protection.}
  \label{fig:circle_cross}
\end{figure}

\par Each video was prefaced with instruction frames. The seven edited videos were then concatenated into a single $21$-minute (excluding instructions and questions) video per task, resulting in a total of $63$ minutes of experimental data per participant. Because the same seven videos appeared in all three tasks, stimulus familiarity could accumulate across tasks. Therefore, both the task order and the video order within each task were randomized across participants to distribute any such effect across the three conditions rather than confounding a specific contrast. A full list of the task and video orders for each participant is provided in the Appendix (Table \ref{tab:order_full}).

\subsection{Tasks}
\label{sec:tasks}

\par
\begin{figure*}[!t]
  \centering
  \includegraphics[width=0.7\textwidth]{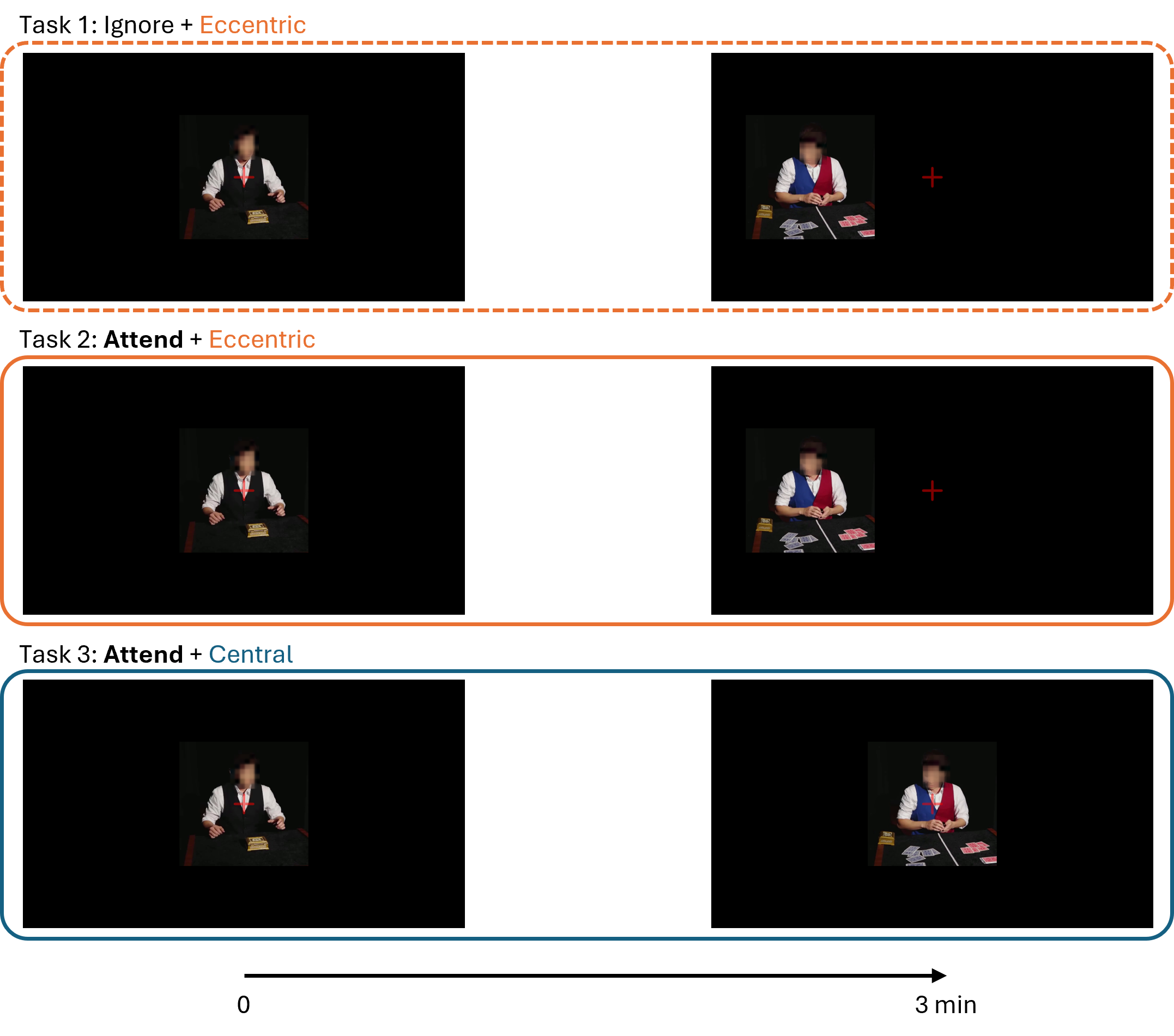}
  \caption{Illustration of the three experimental tasks. The faces are blurred for identification protection. In Task 1, participants ignored the videos while fixating centrally as the videos moved to the periphery. In Task 2, they attended to the videos while fixating centrally as the videos moved to the periphery. In Task 3, they attended to the videos while fixating centrally with the videos remaining in the center. Note that the foreground video does not always have a black or dark background as shown in this example.}
  \label{fig:tasks}
\end{figure*}

 Participants were instructed to maintain fixation on the central cross throughout all tasks. In Task 1 (ignore condition), they were instructed to ignore the video content. To discourage attending to the video, they were also instructed to press the “+” key with their preferred hand on a keyboard whenever the fixation cross dimmed. In Tasks 2 and 3 (attend conditions), participants attended to the videos while continuing to fixate. Because object eccentricity varies in Task 2 but remains constant in Task 3, the latter serves as a control to isolate the effects of eccentricity. To maintain consistent engagement with the video content, participants performed a vigilance task requiring them to press the “o” key with their preferred hand whenever a subtle red circle appeared within the video. The circle was kept deliberately subtle to minimize the chance of detecting it without attending to the video content, while a more salient target could also have competed with it. Both detection tasks served only to sustain engagement, and the key presses were not logged. The rationale and its limitations are discussed in Section~\ref{sec:manip_check}. An illustration of the attention and eccentricity conditions across the three tasks is shown in Figure \ref{fig:tasks}, and a summary of the tasks is provided in Table \ref{tab:tasks_summary}.

\begin{table*}[htbp]
    \centering
    \caption{Summary of experimental tasks.}
    \begin{tabular}{lccccc}
    \toprule
Task & Attention & Eccentricity & Gaze & Response trigger & Response \\
\midrule
Task 1 & Ignore & Central to peripheral & Fixed & Fixation cross dimming & Press “+” key \\
Task 2 & Attend & Central to peripheral & Fixed & Red circle appearance & Press “o” key \\
Task 3 & Attend & Central & Fixed & Red circle appearance & Press “o” key \\
    \bottomrule
    \end{tabular}
    \label{tab:tasks_summary}
\end{table*}

\subsection{Data acquisition}
\label{sec:data_acquisition}

\par Stimuli were presented on a $22$-inch monitor ($1920\times 1080$, \SI{60}{\hertz} refresh rate). Participants sat approximately \SI{130}{\centi \meter} from the screen and were instructed to maintain a stable posture and minimize head and body movements during video playback.

\par EEG data were recorded using a $64$-channel BioSemi ActiveTwo system at \SI{2048}{\hertz}. Four electrooculogram (EOG) electrodes were placed above/below the right eye and at the outer canthi of both eyes. During data acquisition, signal quality was monitored in real time via visual inspection of the signal traces. Channels were manually logged as bad if they displayed clear, irremediable artifacts. These included excessive high-frequency noise, severe signal drifts (often indicating unstable electrode contact), or flat signals typically resulting from conductive gel bridges between adjacent electrodes. Video onset was synchronized with EEG using a photodiode placed at the top-right corner of the screen, which detected a black-to-white box transition at the start of each video. This area was covered with black tape to avoid distracting the participants.

\par Gaze data were recorded using the Pupil Labs NEON eye tracker (\SI{200}{\hertz}). A world-facing camera recorded the screen content from the participant's perspective. Gaze-video synchronization was achieved using a QR code embedded in the instruction frames. The time point when the QR code disappeared in the world-facing camera recording was marked as video onset, and gaze timestamps were aligned to this reference point. Four reference markers were placed around the physical monitor. The Pupil Cloud Marker Mapper algorithm, which uses the world-facing camera to continuously detect these markers and define a static 2D surface corresponding to the screen, allowed us to export normalized, screen-relative gaze positions that are robust to head movements.

\section{Methods}
\label{sec:method}

\subsection{Preprocessing}
\label{sec:prepro}

\par The EEG and EOG data were preprocessed using standard techniques as in Yao et al. \cite{yao2024identifying}: segmentation based on photodiode synchronization signals, bad channel interpolation, average referencing, high-pass filtering (\SI{0.5}{\hertz} cutoff), and notch filtering (\SI{50}{\hertz}). The EOG data were then linearly regressed out from EEG to suppress ocular artifacts. All data were downsampled to \SI{30}{\hertz} with anti-aliasing to match the video frame rate. Rather than using raw video data, we extracted the object-based optical flow, which is a feature that quantifies the overall motion of a single object\footnote{The videos in our experiment only contain one object, namely the human performer.} in the video at each moment in time, resulting in a 1D time series with the same length as the original video \cite{yao2024identifying}. The EEG signals and video features were centered to have zero mean per video.

\par
\begin{figure}[htbp]
    \centering
    \includegraphics[width=0.8\linewidth]{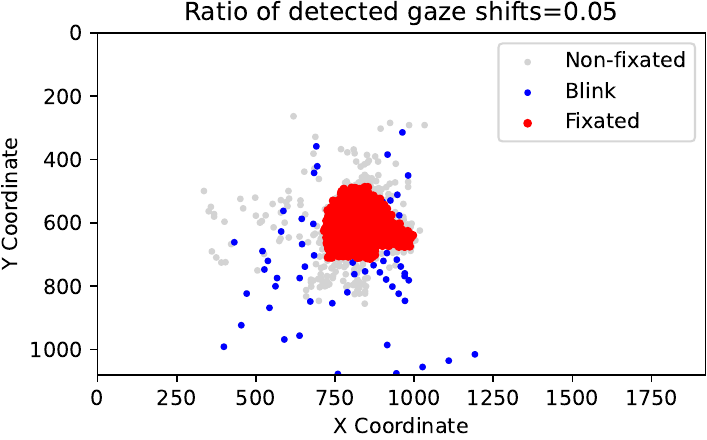}
    \caption{Illustration of non-fixation detection for a representative participant (subject 4) during a video in Task 1. Dots indicate gaze positions with colors representing fixations (red), eye blinks (blue), and non-fixations (gray).}
    \label{fig:gaze_example}
\end{figure}

 Gaze coordinates, along with blink and saccade annotations, were exported from the eye tracker to identify non-fixation periods. First, gaze coordinates were clustered with the density-based spatial clustering algorithm DBSCAN \cite{schubert2017dbscan}, which is well suited for identifying dense fixation regions without requiring a predefined number of clusters. DBSCAN was run on the on-screen gaze coordinates (in pixels) with a neighborhood radius eps $= 10$ pixels and a minimum of $5$ samples per core point (minPts $= 5$). The largest cluster, corresponding to the most frequently viewed region, was interpreted as the fixation cluster. To verify the physiological validity of these clusters, we calculated the 95th percentile radius from the cluster centroid, yielding an average fixation radius of approximately 1.22° of visual angle across all tasks (Table \ref{tab:gaze_stats}). This aligns well with standard thresholds for human foveal vision. Samples falling outside this cluster were classified as non-fixations, except when blink annotations indicated that the deviation was caused by an eye blink. Saccade annotations served as a secondary check for non-fixation samples identified by DBSCAN. Any sample annotated as a saccade was categorized as a non-fixation, overriding its cluster assignment. An example of the gaze samples and the resulting fixation cluster for a representative participant in Task 1 is shown in Figure~\ref{fig:gaze_example}. The mean and standard deviation of the proportion of detected gaze shifts across participants and videos for each task are summarized in Table~\ref{tab:gaze_stats}.

\begin{table*}[t]
    \centering
    \caption{Gaze shift and fixation cluster statistics across participants and videos per task. The table presents the mean and standard deviation for the percentage of samples classified as gaze shifts, as well as the radius of the primary fixation cluster. The fixation radius is expressed in Degrees of Visual Angle (DVA).}
    \begin{tabular}{lcccc}
        \toprule
        & \multicolumn{2}{c}{\shortstack{Percentage of detected \\ gaze shift samples}} & \multicolumn{2}{c}{\shortstack{Fixation cluster \\ radius (DVA)}} \\
        \cmidrule(lr){2-3} \cmidrule(lr){4-5}
        Task & Mean (\%) & Std (\%) & Mean ($^\circ$) & Std ($^\circ$) \\
        \midrule
        Task 1 (ignore, eccentric) & 3.7 & 3.3 & 1.21 & 0.32 \\
        Task 2 (attend, eccentric) & 5.9 & 5.6 & 1.23 & 0.31 \\
        Task 3 (attend, central)   & 3.6 & 3.5 & 1.21 & 0.33 \\
        \bottomrule
    \end{tabular}
    \label{tab:gaze_stats}
\end{table*}

% \begin{table}[htbp]
%     \centering
%     \caption{Gaze shift statistics across participants and videos per task.}
%     \begin{tabular}{lcc}
%     \toprule
%     Task & Mean (\%) & Std (\%) \\
%     \midrule
%     Task 1 (ignore, eccentric) & 3.7 & 3.3 \\
%     Task 2 (attend, eccentric) & 5.9 & 5.6 \\
%     Task 3 (attend, central) & 3.6 & 3.5 \\
%     \bottomrule
%     \end{tabular}
%     \label{tab:gaze_stats}
% \end{table}
From Table \ref{tab:gaze_stats}, we conclude that the overall percentage of non-fixation samples is low, with Task 2 exhibiting a slightly higher proportion of gaze shifts. This increase may be attributable to the greater difficulty of maintaining central fixation when the attended object moves toward the periphery.

\subsection{Correlation analysis}
\label{sec:corr_analysis}

\par Two types of correlation analyses were performed: (1) stimulus-response correlation between EEG and video stimuli within each individual participant using canonical correlation analysis (CCA) \cite{yao2024identifying}, and (2) inter-subject correlation of EEG responses across participants using correlated component analysis (CorrCA) \cite{dmochowski2012correlated}. For completeness, we briefly summarize both methods here.

\par Denote the EEG signals as $\mathbf X \in \mathbb R^{T\times D_x}$ and video stimuli as $\mathbf Y \in \mathbb R^{T\times D_y}$, where $T$ is the number of samples, $D_x$ is the number of EEG channels, and $D_y$ is the dimension of the video features. In our case $D_x=64$ as we collected $64$-channel EEG, and $D_y=1$ as we used the $1$-dimensional object-based optical flow feature. However, these dimensions were expanded by adding time-lagged copies of the signals, enabling a data-driven temporal alignment between the two modalities and realizing a data-driven finite impulse response filtering that suppresses irrelevant frequency bands. Specifically, including $L_x-1$ time lags together with an offset $\Delta_x$ that accounts for latencies between the EEG and the stimulus signal for an EEG sample $\mathbf x(t)=[x_1(t),...,x_{D_x}(t)]\in \mathbb R^{1 \times D_x}$ yields the extended vector:

\begin{equation}
  \begin{aligned}
    \tilde{\mathbf x}(t) = [&x_1(t+\Delta_x), x_1(t-1+\Delta_x), \ldots, \\& x_1(t-L_x+1+\Delta_x), \ldots, x_{D_x}(t+\Delta_x),\\& x_{D_x}(t-1+\Delta_x), \ldots,   x_{D_x}(t-L_x+1+\Delta_x) ] \\& \in \mathbb R^{1 \times D_xL_x}.
  \end{aligned}
\end{equation}
The inclusion of this fixed offset $\Delta_x$ is particularly useful to compensate for a delay in the neural response to the video while avoiding the introduction of a large number of time lags, which would unnecessarily expand the dimension of $\mathbf X$ and $\mathbf Y$. A similar operation, with $L_y-1$ time lags and offset $\Delta_y$, was applied to the video features. After this extension, the dimensions of $\mathbf X$ and $\mathbf Y$ became $T\times D_x L_x$ and $T\times D_y L_y$, respectively.

\par Recall that the signals were centered to have zero mean. The CCA model finds linear maps $\mathbf w_x$ and $\mathbf w_y$ such that the correlation between the transformed signals is maximized \cite{hotelling1992relations}:
\begin{equation}
  \begin{aligned}
  & \underset{\mathbf{w}_x, \mathbf{w}_y}{\text{maximize}}
  & & \transpose{\mathbf{w}}_x\transpose{\mathbf{X}}\mathbf Y\mathbf{w}_y \\
  & \text{subject to}
  & & \transpose{\mathbf{w}}_x\transpose{\mathbf{X}}\mathbf X\mathbf{w}_x = 1, \\
  &&& \transpose{\mathbf{w}}_y\transpose{\mathbf{Y}}\mathbf Y\mathbf{w}_y = 1.
  \end{aligned}
  \label{eq:cca}
  \end{equation}
The optimized $\mathbf w_x$ and $\mathbf w_y$ are the first canonical components, yielding the first canonical directions $\mathbf X \mathbf w_x$ and $\mathbf Y \mathbf w_y$. A higher correlation between these canonical directions indicates stronger neural tracking of the video features.

Higher-order components can be obtained by solving the same optimization problem with an additional constraint that these higher-order canonical directions are orthogonal to all previous canonical directions. Aggregating canonical components from first to $K$-th order as $\mathbf W_x\in \mathbb R^{D_x L_x\times K}$, $\mathbf W_y\in \mathbb R^{D_yL_y \times K}$, the solution can be obtained by solving a generalized eigenvalue decomposition (GEVD) problem:
\begin{equation}
\begin{aligned}
    \mathbf R_{xy}\mathbf R_{yy}^{-1}\mathbf R_{yx} \mathbf W_x &= \mathbf R_{xx}\mathbf W_x \boldsymbol{\Lambda}, \\
    \mathbf R_{yx}\mathbf R_{xx}^{-1}\mathbf R_{xy} \mathbf W_y &= \mathbf R_{yy}\mathbf W_y \boldsymbol{\Lambda},
\end{aligned}
\label{eq:cca_gevd}
\end{equation}
where $\mathbf R_{xy} = \transpose{\mathbf X}\mathbf Y$, $\mathbf R_{yx} = \transpose{\mathbf Y}\mathbf X$, $\mathbf R_{xx} = \transpose{\mathbf X}\mathbf X$, $\mathbf R_{yy} = \transpose{\mathbf Y}\mathbf Y$, and $\boldsymbol{\Lambda}$ is a diagonal matrix with eigenvalues in descending order.

\par CorrCA extends CCA to a group-level analysis, enabling the extraction of common neural components from the EEG data of multiple (more than two) participants. Since the EEG signals of all participants are time-locked on the same videos, these common components are expected to be neural responses to these videos \cite{dmochowski2012correlated}. Using subscript $n$ to denote the subject index, EEG signals from different subjects are denoted as $\mathbf X_n$, where $n = 1, ..., N$. CorrCA finds a shared\footnote{Other generalizations of CCA exist, which allow using a different spatial filter for each participant \cite{kettenring1971canonical, geirnaert2023stimulusInformed}. We opted for CorrCA here because of the limited dataset size, since the reduction in degrees of freedom (i.e., using the same $\mathbf W_s$ for all participants) has a regularizing effect on the transformation \cite{geirnaert2023stimulusInformed}.} spatial filter $\mathbf W_s$ that maximizes the approximate average pairwise correlations between all subject pairs: 
\begin{equation}
  \begin{aligned}
  & \underset{\mathbf{W}_s}{\text{maximize}}
  & & \sum_{i=1, i\neq j}^N\sum_{j=1}^N\trace{\transpose{\mathbf W}_s \mathbf R_{ij}\mathbf W_s} \\
  & \text{subject to}
  & & \sum_{i=1}^N\transpose{\mathbf W}_s\mathbf R_{ii}  \mathbf W_s = \mathbf I_K,
  \end{aligned}
  \label{eq:corrca_multi}
  \end{equation}
where $\trace{\cdot}$ denotes the trace, $\mathbf W_s$ contains the first $K$ canonical components, and $\mathbf R_{ij} = \transpose{\mathbf X}_i\mathbf X_j$ is the cross-covariance matrix between participants $i$ and $j$. This optimization problem can also be solved via a GEVD:
\begin{equation}
  \label{eq:corrca_gevd}
  \left(\sum_{i=1}^N \sum_{j=1}^N \mathbf R_{ij}\right) \mathbf W_s = \left(\sum_{i=1}^N \mathbf R_{ii}\right) \mathbf W_s \mathbf \Lambda.
\end{equation}
The eigenvalues in $\boldsymbol{\Lambda}$ are in descending order. Inter-subject correlation (ISC) for the $k$-th component is defined as the average correlation between all subject pairs:
\begin{equation}
\text{ISC}_k = \frac{2}{N(N-1)}\sum_{i=1}^{N}\sum_{j=i+1}^{N} \frac{\transpose{\mathbf w}_{s,k} \mathbf R_{ij}\mathbf w_{s,k}}{\sqrt{\transpose{\mathbf w}_{s,k} \mathbf R_{ii} \mathbf w_{s,k}}\sqrt{\transpose{\mathbf w}_{s,k} \mathbf R_{jj} \mathbf w_{s,k}}},
\end{equation}
where $\mathbf w_{s,k}$ is the $k$-th column of $\mathbf W_s$. Higher ISC indicates stronger neural synchronization across participants, which has been linked to higher attention levels \cite{dmochowski2012correlated, ki2016attention}.

To interpret the weights of the EEG encoders ($\mathbf W_x$ in CCA and $\mathbf W_s$ in CorrCA), we compute the forward model, which maps the extracted components to the original EEG space \cite{haufe2014interpretation}. The weights in the forward model reveal how strongly the extracted components are reflected in each EEG channel. The forward models for CCA ($\mathbf F_x$) and CorrCA ($\mathbf F_s$) are defined as
\begin{equation}
  \label{eq:forward_model}
  \min _{\mathbf F_x} \|\mathbf X-\mathbf X\mathbf W_x \transpose{\mathbf F}_x \|_{\mathrm F}^2
\end{equation} 
and 
\begin{equation}
  \label{eq:forward_model_avg}
  \min_{\mathbf F_s} \left\|\begin{bmatrix}
    \mathbf X_1 \\ \vdots \\ \mathbf X_N
  \end{bmatrix} -
  \begin{bmatrix}
    \mathbf X_1 \mathbf W_s \\ \vdots \\ \mathbf X_N \mathbf W_s
  \end{bmatrix}\transpose{\mathbf F}_s \right\|_{\mathrm F}^2,
\end{equation}
respectively. The solutions to these least-squares problems are given by:
\begin{equation}
\mathbf F_x = \mathbf R_{xx}\mathbf W_x (\transpose{\mathbf W}_x \mathbf R_{xx}\mathbf W_x)^{-1},
\label{eq:cca_forward_model}
\end{equation}
\begin{equation}
  \mathbf F_s = \sum_{i=1}^N \mathbf R_{ii}\mathbf W_s(\sum_{i=1}^N \transpose{\mathbf W}_s \mathbf R_{ii}\mathbf W_s)^{-1}.
\label{eq:corrca_forward_model}
\end{equation}
To prevent our spatial topographies from reflecting overfitted noise in the training set, the covariance matrices $\mathbf{R}_{xx}$ and $\mathbf{R}_{ii}$ are computed using unseen test data.

\subsection{Parameters and evaluation}
\label{sec:params_eval}

\par Temporal lag parameters were set according to Yao et al. \cite{yao2024identifying}. For CCA, we used $L_x = 3$ with $\Delta_x = 1$, capturing EEG information from \SI{-33}{\milli\second} to \SI{33}{\milli\second} around the current time point, and $L_y = 15$ with $\Delta_y = 0$, capturing video features from approximately 500 ms of history to the current time. For CorrCA, $L_x = 5$ and $\Delta_x = 2$ were used, covering \SI{-67}{\milli\second} to \SI{67}{\milli\second}.

\par We adopted leave-one-video-out cross-validation in all experiments. Unless stated otherwise, training was performed on the combined data from all three experimental tasks (54 minutes per participant: 3 tasks $\times$ 6 videos $\times$ 3 minutes). This was an empirical decision, as aggregating data across all tasks yielded better decoding performance than training on individual tasks alone. A possible explanation is that even in Task 1, where participants were instructed to ignore the video, passive visual processing still engages the same neural pathways, providing useful information for estimating the spatial filters. In the stimulus-response correlation analysis, non-fixation periods identified in Section~\ref{sec:prepro} were excluded from testing, along with a 0.5-second margin selected to outlast saccadic spike potentials (tens of milliseconds \cite{keren2010saccadic}) and post-saccadic lambda waves (100-250~ms \cite{brigo2011lambda}) while preventing severe data loss, to strictly control for eye-movement confounds. Therefore, the size of the test set varied across tasks and participants. For CorrCA, all data were retained because the proportion of non-fixation samples was small and removing non-fixation periods would disrupt temporal alignment across participants. Test segments were defined as 45-second windows sampled from the test set with overlap, randomly drawn with an average interval of 2 seconds. The two analyses also treat the marker events of Section~\ref{sec:protocol} differently. CorrCA extracts activity that is correlated across participants without knowledge of the stimulus, and because the event timing was identical for all participants, the responses these markers evoke are synchronized across subjects and could inflate the ISC. The surrounding intervals are therefore excluded (Section~\ref{sec:group_analysis}). CCA instead extracts the component correlated with the object-based optical flow. The event onsets were placed at random with respect to the video content, so the activity they evoke is uncorrelated with that feature and is not extracted by the canonical filters. Therefore, there is no need to discard the corresponding intervals in the match-mismatch analysis.

\par For CCA, instead of using raw correlation coefficients, which are rather noisy, we performed match-mismatch (MM) decoding to quantify stimulus-response coupling strength \cite{de2021auditory}. Given an EEG segment, the task is to identify the corresponding video segment from two candidates: the correct match and a mismatch (a segment from the same video but different time point). The procedure applies the trained EEG decoder $\mathbf{w}_x$ and video encoder $\mathbf{w}_y$ to compute canonical correlations between the test EEG segment and both candidate video segments. The candidate with the higher sum of the first two canonical correlation components was selected as the predicted match. This decision metric was made a priori based on previous studies in neural tracking of natural videos using the same CCA-based methodology \cite{yao2024identifying,yao2025eeg}. Decoding accuracy was then calculated across all test trials. To assess whether decoding performance was significantly above chance, we circularly permuted the sequence of video segments such that the temporal alignment between EEG and video segments was disrupted and both MM candidates became mismatches (100 permutations per participant). These per-participant permutation results were pooled across all 14 participants to construct a single group-level null distribution comprising 1400 total permutations. The significance bound was set at the $97.5$th percentile of this null distribution. Circular permutation was selected for the MM-decoding null because it preserves the marginal statistics and within-segment temporal structure of both the EEG and the video features while destroying only their cross-modal temporal correspondence. Since the mismatch segment was drawn at random for each trial, the resulting decoding accuracies could vary slightly between runs. To obtain stable estimates for the between-task and between-protocol comparisons in Section~\ref{sec:results}, we repeated the full decoding procedure five times and averaged the per-participant accuracies across runs before applying the Wilcoxon signed-rank and Mann-Whitney U tests. 

\par For CorrCA, ISC values were computed and averaged across all test segments. Significance was assessed using a permutation test analogous to the MM-decoding null. For each participant, the EEG time series was circularly shifted by a random offset to disrupt the cross-subject temporal alignment while preserving the temporal structure and marginal statistics within each participant's data. ISC was then recomputed on these permuted data (500 permutations per fold, yielding 3500 total permutations), and the significance threshold was set at the 97.5th percentile of the resulting null distribution, corresponding to a two-sided test with $\alpha = 0.05$.

\section{Results}
\label{sec:results}

\subsection{Effect of eccentricity on decoding performance}
\label{sec:tasks_comparison}

\par
\begin{figure}[htbp]
    \centering
    \includegraphics[width=0.65\linewidth]{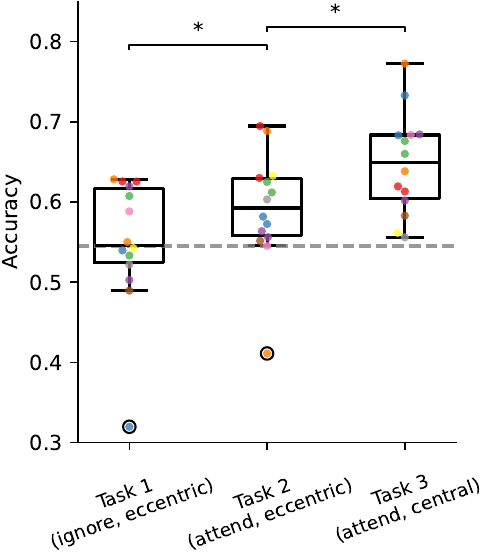}
    \caption{Match-mismatch (MM) decoding accuracy across three experimental tasks on 45-second test segments. Individual participant accuracies are shown as colored dots, with statistical outliers circled (identified by the Tukey $1.5\times$IQR rule; outliers were retained in all Wilcoxon signed-rank tests). Box plots display median and interquartile range, with whiskers extending to extreme values. The dashed line indicates the significance bound for above-chance performance, defined as the 97.5th percentile of the circular permutation null distribution (1400 permutations; see Section~\ref{sec:params_eval}). Asterisks denote significant between-task differences: Task 1 $<$ Task 2 ($p=0.032$, $Z=1.85$, rank-biserial $r=0.56$; corrected $p_{\text{adj}}=0.048$), Task 2 $<$ Task 3 ($p = 0.024$, $Z=1.98$, rank-biserial $r=0.60$; corrected $p_{\text{adj}}=0.048$), demonstrating both attentional effects and eccentricity confounds.}
    \label{fig:acc_tasks}
\end{figure}

 Tasks 1 and 2 both had videos gradually moving from center to periphery, differing only in attention condition: participants ignored video content in Task 1 but attended covertly in Task 2. Comparing these two tasks reveals whether attentional modulation persists when the visual objects are not centrally located. Task 3 maintained videos at the center throughout while the participants attended to the content. Comparing Task 3 with Task 2 isolates the effect of eccentricity on decoding performance. MM accuracies for all three tasks are shown in Figure \ref{fig:acc_tasks}.

\par In Task 1, half of the participants (7 out of 14) failed to achieve above-chance decoding accuracy, with a median accuracy of $0.55$. Tasks 2 and 3 yielded median accuracies of $0.59$ and $0.65$, respectively, with most participants achieving significantly above-chance performance. To evaluate the distinct effects of attention and eccentricity, we conducted two comparisons using Wilcoxon signed-rank tests. Because both contrasts tested directional hypotheses specified a priori (Task 1 $<$ Task 2 and Task 2 $<$ Task 3), one-sided tests were used, and the two resulting $p$-values were corrected for multiple comparisons with the Holm-Bonferroni procedure. The Task 1 accuracy was significantly lower than that of Task 2 ($p=0.032$, $Z=1.85$, rank-biserial $r=0.56$; corrected $p_{\text{adj}}=0.048$), confirming that attentional modulation persists even when objects move away from the visual field center. Additionally, Task 2 accuracy was significantly lower than that of Task 3 ($p = 0.024$, $Z=1.98$, rank-biserial $r=0.60$; corrected $p_{\text{adj}}=0.048$), demonstrating that shifting the visual stimulus to peripheral locations significantly impairs decoding performance even under identical attentional conditions. This provides direct evidence that eccentricity acts as a confounding factor in visual attention decoding, when neural tracking strength is used to index visual attention. To ensure that this effect is not an artifact of the filters representing a compromise fit between the central and eccentric conditions, we repeated the analysis with filters estimated separately within each task and confirmed that the eccentricity effect persisted (see Appendix \ref{sec:app_within_task}). Although the corrected $p$-values lie close to the conventional $0.05$ significance threshold, both contrasts remained significant after correction, and the effect sizes (rank-biserial $r=0.56$ and $0.60$) indicate that these are substantial rather than marginal effects. Moreover, we note that this reduction was measured in the neural tracking of object motion. Our design therefore cannot determine whether it is specific to motion processing or reflects a broader decrease in visual stimulus-response coupling at the periphery.

\subsection{Gaze fixation versus free viewing}
\label{sec:protocols_comparison}

\par
\begin{figure}[htbp]
    \centering
    \includegraphics[width=0.65\linewidth]{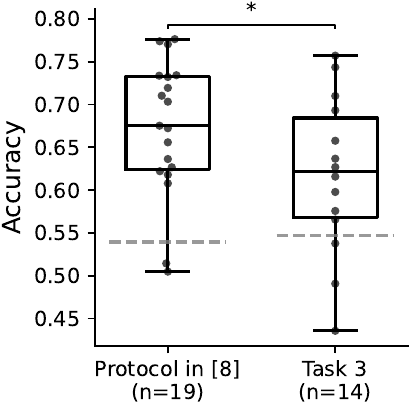}
    \caption{Comparison of match-mismatch (MM) decoding accuracy on 45-second test segments under free-viewing (19 participants, \cite{yao2024identifying, yao_2024_10512414}) and gaze-fixation (14 participants) protocols. Dots represent the accuracies of individual participants. Box plots display median and interquartile range, with whiskers extending to extreme values. The dashed lines indicate the significance bound for above-chance performance for each protocol, defined as the 97.5th percentile of the circular-permutation null distribution (calculated using 100 permutations per participant, yielding 1900 total permutations for the protocol in \cite{yao2024identifying} and 1400 for the current gaze-fixation protocol; see Section~\ref{sec:params_eval}). Gaze fixation significantly reduced decoding accuracy compared to free viewing (Mann-Whitney U test, $p = 0.042$, $U=181.0$, rank-biserial $r=0.36$).}
    \label{fig:acc_protocols}
\end{figure}

 In this section, we compared the decoding performance under our gaze-fixation protocol with previous free-viewing studies. Using the free-viewing dataset from Yao et al. \cite{yao2024identifying}\footnote{The original study included 20 subjects, with 19 consenting to make their data publicly available in \cite{yao_2024_10512414}.}, we selected only the data corresponding to the same video stimuli used in our current study, yielding 21 minutes per participant (for both datasets). EOG data were linearly regressed out from EEG during preprocessing, as in the present dataset. For fair comparison, only the data from Task 3 (attention condition with object in the center) were used from the current study in both training and testing to have an equal amount of test and train data, and to remove effects of eccentricity. The same leave-one-video-out cross-validation was performed. Results are presented in Figure \ref{fig:acc_protocols}.

\par The gaze-fixation protocol yielded a mean accuracy of $0.62$, compared to $0.67$ for free viewing. To test whether this reduction was significant, we used a Mann-Whitney U test since the participant groups differed between the two protocols. The p-value was $0.042$ ($U=181.0$, rank-biserial $r=0.36$), suggesting significant performance loss under gaze fixation. We note that the results in Figure \ref{fig:acc_protocols} differ from the Task 3 results shown in Figure \ref{fig:acc_tasks}, which is due to the difference in the amount of training data in both analyses (the results in Figure \ref{fig:acc_tasks} use 3 times more training data, as training data is aggregated across the 3 tasks). 

\subsection{Effect on group analysis}
\label{sec:group_analysis}

\par
\begin{figure}[htbp]
    \centering
    \includegraphics[width=\linewidth]{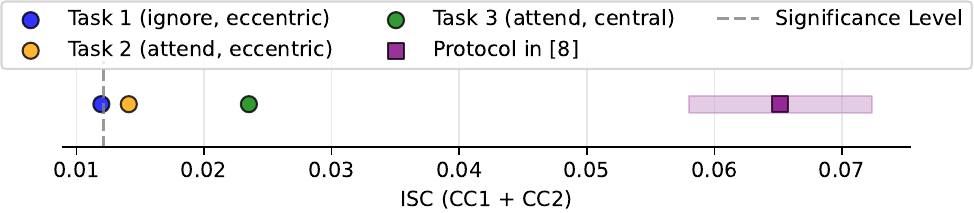}
    \caption{Sum of the inter-subject correlations (ISCs) of the first two canonical components across experimental conditions. ISC values are computed separately for each task under the gaze-fixation protocol and compared with the free-viewing protocol. To match the participant count of the current study, the free-viewing reference is generated by repeatedly drawing 14 participants at random from the original dataset \cite{yao2024identifying, yao_2024_10512414} (5 independent draws) and computing the ISC for each draw. The shaded region represents the standard deviation across these draws and the square represents the mean. The vertical dashed line marks the significance threshold, defined as the 97.5th percentile of the circular permutation null distribution (500 permutations per fold, totaling 3500; see Section~\ref{sec:params_eval}).}
    \label{fig:isc}
\end{figure}

 The previous sections investigated the stimulus-response coupling on a per-subject basis. In this section, we performed a group analysis based on the ISC to investigate the coherence of the neural responses across participants, without explicitly linking these to the stimulus signal. We computed ISCs for all three tasks in the current protocol, excluding time points from \SI{-250}{\milli\second} to \SI{500}{\milli\second} relative to the onset of the artificially added markers (fading of the fixation cross and red circle), irrespective of whether the task required a response to them, to avoid contamination from event-evoked and motor responses that are time-locked and therefore synchronized across participants. The results are shown in Figure \ref{fig:isc}, alongside the free-viewing protocol results from \cite{yao2024identifying} for comparison. The free-viewing results were computed by randomly selecting 14 participants (repeated 5 times) from the original dataset \cite{yao2024identifying, yao_2024_10512414} to match the current study's participant count, as the number of participants influences the ISC values. As in Section~\ref{sec:protocols_comparison}, only data corresponding to the same video stimuli as in the current gaze-fixation protocol were used, yielding $18 \times 14$ minutes of training data per fold. Note that this was only one-third of the training data available in the gaze-fixation protocol, where data from all three tasks were aggregated for training.

\par The ISC pattern mirrored the decoding accuracy trends from Section \ref{sec:tasks_comparison}: Task 1 showed the lowest ISC (slightly below significance threshold), followed by Task 2, with Task 3 exhibiting the highest ISC. This confirmed both attention- and eccentricity-related attenuation of the stimulus-evoked neural responses, reducing the ISC across participants. However, Task 2 ISC was only marginally above significance, showing a larger gap with Task 3 than with Task 1. This suggests that eccentricity may have an even stronger impact on ISC than attention effects. Additionally, free-viewing protocol results were substantially higher despite a smaller training data size, which aligns with the different decoding performance observed in Section \ref{sec:protocols_comparison}. 

\subsection{Spatial Topographies of the Decoding Models}
\label{sec:spatial_topographies}

\par
\begin{figure}[!t]
  \centering
  \begin{subfigure}{0.5\textwidth}
    \centering
    \includegraphics[width=0.9\linewidth]{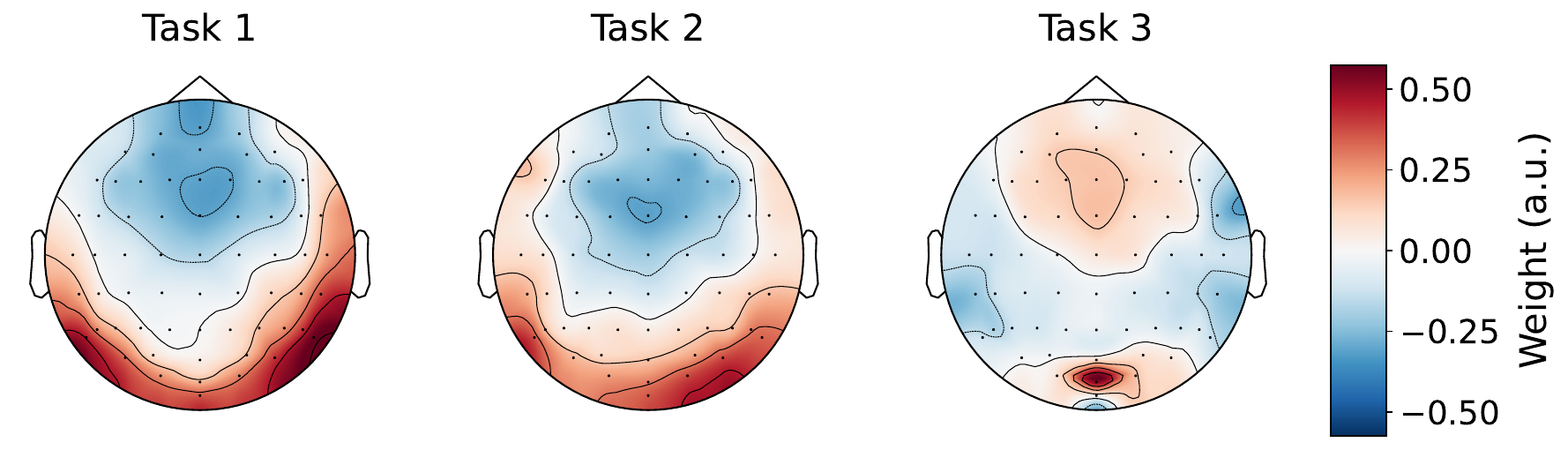} % upper plot
    \caption{}
    \label{fig:cca_fm}
  \end{subfigure}

  \par\medskip
  \begin{subfigure}{0.5\textwidth}
    \centering
    \includegraphics[width=0.9\linewidth]{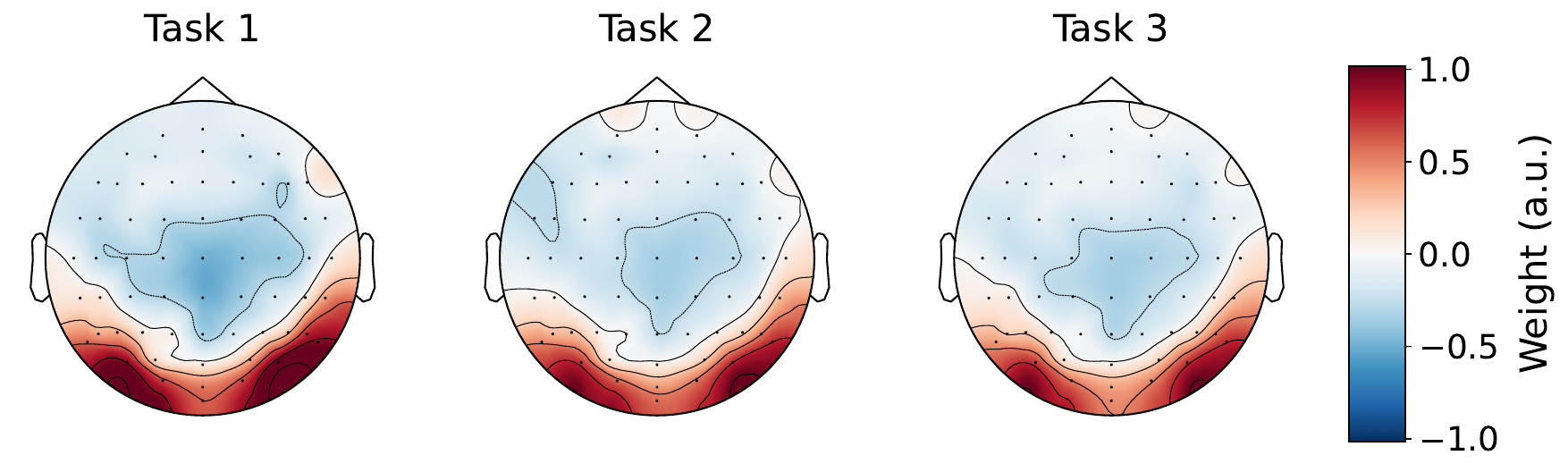} % lower plot
    \caption{}
    \label{fig:corrca_fm}
  \end{subfigure}

  \caption{Spatial topographies (forward models) of the decoding models for each task. (a) CCA: forward model of the most strongly correlated canonical component, averaged across participants. (b) CorrCA: forward model of the component with the highest ISC. Since the polarity of each component is arbitrary, only the spatial pattern is interpretable.}
  \label{fig:forward_models}
\end{figure}

 To gain insights into the neural sources contributing to the decoding performance and ISC, we computed the forward models for both CCA and CorrCA components using \eqref{eq:cca_forward_model} and \eqref{eq:corrca_forward_model}, respectively. To avoid introducing noise, only the forward models corresponding to significant results across folds were retained and averaged. For CCA, we averaged the forward models across all participants to generate the spatial topography of the first component, which exhibited the largest stimulus-response correlation. For CorrCA, we visualized the forward model of the group decoder that yielded the highest ISC. Because canonical components have an arbitrary polarity, the per-participant forward models were sign-aligned prior to averaging. Therefore, only the spatial weighting pattern and not its absolute polarity should be interpreted. The resulting spatial topographies for the different tasks are shown in Figure \ref{fig:forward_models}.

\par Across all topographies, we observed no strong activation in the frontal channels (relative to other regions), indicating that eye-movement artifacts, which were at least partially removed using EOG regression (see Section \ref{sec:prepro}), probably contributed minimally to decoding performance and ISC. Beyond this shared absence of frontal involvement, the topographies differed systematically across tasks and between the two decoding methods. We interpret these patterns and their link to the eccentricity effect in Section~\ref{sec:ecc_reorg_topo}.

\section{Discussion}
\label{sec:discussion}

\subsection{Group analysis is more susceptible to eye movement artifacts than stimulus-response analysis}
\label{sec:GA_SRA_eye}

\par
\begin{figure}[!t]
  \centering
  \begin{subfigure}{0.5\textwidth}
    \centering
    \includegraphics[width=0.9\linewidth]{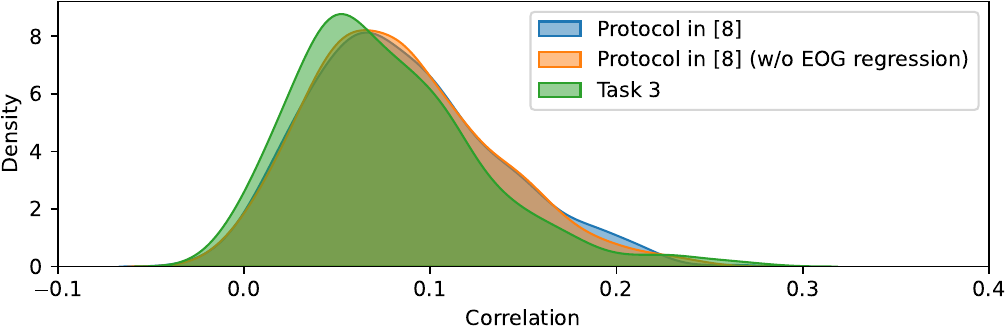} % upper plot
    \caption{}
    \label{fig:dist_corr}
  \end{subfigure}

  \par\medskip
  \begin{subfigure}{0.5\textwidth}
    \centering
    \includegraphics[width=0.9\linewidth]{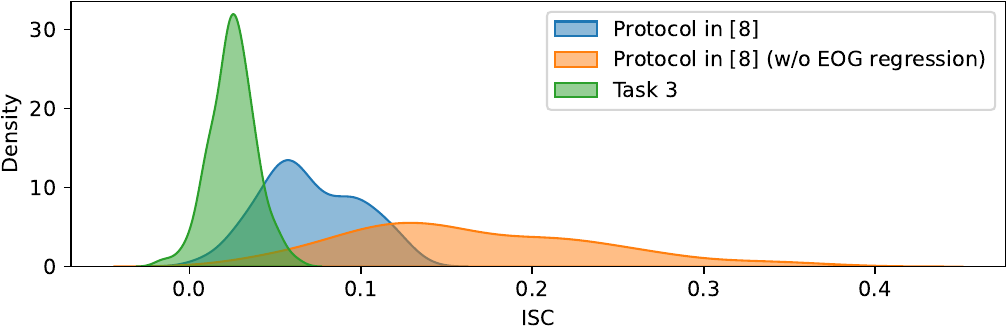} % lower plot
    \caption{}
    \label{fig:dist_isc}
  \end{subfigure}

  \caption{Distributions of (a) stimulus-response correlations and (b) inter-subject correlations across participants for 45-second segments under the free-viewing and gaze-fixation protocols. The values represent the sum of the first two components. For comparison, results obtained from EEG without EOG regression are also included. For the gaze-fixation condition, only Task 3 (attend, central) is shown. Curves are Gaussian kernel density estimates computed from approximately 500 45-second segments (pooled across participants) for (a), and 100 45-second segments for (b). The bandwidth was selected automatically by Scott's rule, the default of seaborn's \texttt{kdeplot}.}
  \label{fig:dist_2x1}
\end{figure}

 From Figure \ref{fig:isc}, we observed that the decrease in ISC from free-viewing to gaze-fixation protocols appeared more pronounced than the corresponding decrease in decoding accuracy shown in Figure \ref{fig:acc_protocols}. However, note that Figure \ref{fig:acc_protocols} displays MM accuracy, whereas ISC is a correlation metric. To more directly compare the effects of gaze fixation on group-level and individual-level analyses, we plotted the distributions of stimulus-response correlations (aggregated across participants) and ISCs in Figure \ref{fig:dist_2x1} for both the free-viewing protocol and the gaze-fixation protocol (Task 3: attend, central). For reference, we also included results obtained without EOG regression under the free-viewing protocol.

\par For the individual stimulus-response analysis, the distributions with and without EOG regression under the free-viewing protocol overlapped substantially. Under the gaze-fixation protocol (with EOG regression), the distribution shifted slightly leftward with a thinner right tail, indicating that gaze fixation reduced stimulus-response correlations rather mildly. In contrast, the group analysis showed a much stronger effect. After regressing out EOG, the free-viewing ISC distribution already exhibited a pronounced leftward shift, indicating that EOG-capturable ocular artifacts contributed substantially to ISC. Under the gaze-fixation protocol, ISC values decreased further and the distribution became narrower. This additional reduction beyond EOG regression may be partially attributed to the absence of neural activities such as motor signals from saccade planning and execution that are not captured by EOG channels but are nonetheless synchronized across participants viewing the same dynamic content. Moreover, standard EOG regression is unable to remove saccadic spike potentials, lambda waves, and gaze-induced visual motion, all of which act coherently across participants to inflate ISC under free-viewing conditions. Together, these results suggest that group analysis is more susceptible to eye-movement confounds than individual stimulus-response analysis.

\par One likely reason is that eye movements tend to be synchronized across participants when viewing the same dynamic visual stimuli, artificially inflating ISC. Stimulus-response analysis, on the other hand, focuses on the relationship between the stimulus and each participant's neural response. Although eye-movement-related activity in the EEG can also contribute to stimulus-response correlations when using motion-based video features, its impact is less pronounced, likely because the relationship between the stimulus and eye movements is indirect and more difficult to capture.

% \par Moreover, in \cite{yao2024identifying}, it was shown that the object-based optical flow feature only explains a marginal part ($7\%$) of the neural response to the video captured by CorrCA. Compared to the large drop in ISC observed after EOG regression or imposing gaze fixation, this suggests that the neural activity captured by CorrCA is affected even more by eye movement artifacts than by genuine neural responses to the video features.

\subsection{Why does gaze fixation reduce decoding accuracy and ISC?}

\par While gaze fixation clearly reduces decoding accuracy and ISC, as shown in Sections \ref{sec:protocols_comparison} and \ref{sec:group_analysis}, the underlying mechanisms remain unclear. The reduction could arise from the loss of synchronized eye-movement information that aids decoding and boosts ISC in free-viewing paradigms, and/or from the increased difficulty of maintaining sustained attention when gaze is constrained. The covert attention paradigm used here, while necessary to control for eye movement confounds, is less naturalistic than overt attention, where eye movements naturally accompany and facilitate engagement. This reduced ecological validity, combined with increased cognitive demand, may also contribute to weaker neural responses. Indeed, most participants reported that maintaining fixation while attending to moving objects was challenging. However, since the ground truth of a participant's actual attentional state is unavailable, disentangling these two factors is impossible here.

\subsection{Free-viewing studies likely rely on genuine neural responses to object motion}

\par In previous free-viewing studies that explicitly addressed eye-movement confounds \cite{yao2024identifying, yao2025eeg}, several analyses were conducted to demonstrate that the observed stimulus-response correlations and above-chance decoding accuracies were not solely driven by eye movement artifacts. These included regressing out EOG signals, restricting analyses to EEG channels less affected by eye movements, and removing data segments containing saccades. Together, these approaches provide strong evidence that EEG-based decoding of visual attention captures genuine neural responses to object motion. However, since it was never possible to completely rule out all eye-movement confounds in free-viewing paradigms, some uncertainty remained.

\par In the present study, strict gaze fixation was imposed to directly eliminate eye-movement confounds. The fact that significant decoding accuracy and ISC persisted under these conditions provides additional support that previous free-viewing results at least partially reflect genuine neural responses to object motion, although the between-cohort nature of the free-viewing/gaze-fixation comparison precludes quantifying the exact proportion attributable to motion-related neural activity. Future work utilizing a within-subjects design, which evaluates the same participants under both free-viewing and fixation conditions, will be necessary to directly quantify these relative contributions to decoding accuracy.

\subsection{Task and method-dependent variations in spatial topographies}
\label{sec:ecc_reorg_topo}

\par Beyond the shared absence of frontal involvement noted in Section~\ref{sec:spatial_topographies}, the topographies varied with both task and decoding method. For CCA, Tasks 1 and 2 yielded similar topographies with broadly distributed occipital and posterior-temporal weights. This suggests that the same stimulus-driven neural sources were engaged despite differing attention conditions. Task 3, however, showed a different topography with focal occipital weights. This eccentricity-dependent reorganization of the scalp topography aligns with recent and foundational visual evoked potential studies \cite{phangwiwat2024sustained, mohr2024visual, di2002cortical}, which demonstrated that the occipital response is midline-symmetric for foveal stimuli but shifts toward the contralateral hemisphere as the stimulus moves peripherally. Because leftward and rightward video displacements in our paradigm were randomized across videos, averaging the forward models superimposes these oppositely lateralized peripheral sources. This superposition explains why the eccentric Tasks 1 and 2 exhibit broader, more bilaterally distributed occipital and posterior-temporal weights compared to the centrally located Task 3. 

\par In contrast, the CorrCA topographies were highly consistent across all tasks, dominated by broadly distributed occipital-temporal weights. It is important to note that EEG components extracted by CCA were steered to maximize the correlation with the object-based optical flow feature, while CorrCA components were steered to maximize ISC across participants, which inherently captures a broader mixture of shared neural responses. As pointed out in \cite{yao2024identifying}, the object-based optical flow feature used in CCA only explains a small portion of inter-subject covariance in EEG, indicating that the remaining covariance may be dominated by other shared neural responses to the video. The consistent broad activation in the occipital-temporal region across all tasks could reflect engagement of specific networks like those for face and body perception \cite{meeren2013different}, or the extraction of other conceptual information during naturalistic action viewing \cite{dima2026distinct}. Attention and eccentricity may simply scale the magnitude of this shared response, resulting in the consistent topography across tasks.

\subsection{What a behavioral record could and could not have established}
\label{sec:manip_check}

\par The key presses in the event detection tasks were not logged (Section~\ref{sec:tasks}), so neither detection accuracy nor reaction times are available. Two limitations follow: we cannot report whether individual participants complied with the vigilance instructions, and we cannot identify erroneous responses.

\par Such a record would not, however, have indexed attention to the video content, which is the quantity our analyses concern. In Task 1, a participant could report every dimming of the cross while still covertly following the performer throughout. In Tasks 2 and 3, detecting the fading circle shows that the video region was being monitored, but not necessarily that attention was directed to the video content, even though the circle was kept subtle (Section~\ref{sec:tasks}) precisely to prevent this. The subtlety also creates the opposite problem: a circle could be missed by a participant who was fully engaged with the video, and that miss would be misread as inattention. This is why we treated the vigilance tasks as a manipulation rather than a measure.

\par Quantifying it would require a different kind of behavioral record. One option is to probe the video content directly after each block, for instance by asking how many rotations the performer made. Such a task can, evidently, not be used in the ignore condition. Ensuring that the video content is genuinely ignored would instead call for a task more demanding than cross-dimming detection, such as a difficult auditory or arithmetic task, where a high score would indicate that the participant was occupied by that task and had little capacity left for the video. Measures of this kind remain indirect, and establishing a ground truth for the attentional state is inherently difficult.

\par Returning to the behavioral task in this study, other information would also be useful if the key-press record were available. For instance, reaction times to visual events could hint at whether participants built up anticipation of those events. However, because our analyses are not event-based, and the missing record concerns only the marker events, our results and overarching conclusions are not substantially affected. In match-mismatch decoding, the marker events are uncorrelated with the optical-flow feature that drives the decoding (Section~\ref{sec:params_eval}), and in inter-subject correlation analysis, the marker intervals are excluded entirely.

\section{Conclusion}
\label{sec:conclusion}

\par This study revealed a spatial confound in EEG-based visual attention decoding by manipulating stimulus eccentricity under gaze fixation. When objects moved toward peripheral locations, lower decoding accuracies and lower ISCs were observed, demonstrating that peripheral objects are more difficult to decode than central objects. This spatial confound represents an important limitation for current attention decoding approaches that assume coupling strength reflects attention levels alone. Comparison with free-viewing conditions showed that gaze fixation reduced both decoding accuracy and ISC, yet both measures remained statistically significant. The latter suggests that neural tracking of video motion still exists and is modulated by attention under strict gaze fixation, and that previous free-viewing studies also partially reflect genuine neural responses to object motion and not only gaze-induced components such as residual EOG artifacts, gaze-related motor neural activity, or gaze shift-evoked neural responses. While our sample size ($N=14$) is relatively small, the observed attention and eccentricity effects were consistent across both individual-level stimulus-response analysis and group-level inter-subject correlation analysis, lending confidence to the robustness of our findings.

\section{Appendix}

\subsection{Task and video presentation order}
\label{sec:app_order}
\par The task and video presentation order for each participant is shown in Table \ref{tab:order_full}. Task 1 (ignore, eccentric) was presented first for 3 participants, second for 4, and third for 7; Task 2 (attend, eccentric) first for 6, second for 6, and third for 2; and Task 3 (attend, central) first for 5, second for 4, and third for 5. Within each task block, the video orders were well distributed: across the 42 blocks all orderings were unique, each video appeared in early, middle, and late positions roughly equally often (mean serial position $3.6$-$4.5$, against a midpoint of $4.0$), and no video tended systematically to precede any other.

\begin{table*}[htbp]
\centering
\caption{Task-block order and within-block video presentation order for each participant ($N=14$). Within each block, the comma-separated numbers are video IDs.}
\label{tab:order_full}
\small
\setlength{\tabcolsep}{4pt}
\begin{tabular}{llll}
\toprule
Participant & 1st block & 2nd block & 3rd block \\
\midrule
1 & Task 3: 12, 08, 05, 16, 07, 04, 14 & Task 1: 05, 14, 16, 08, 04, 07, 12 & Task 2: 12, 05, 07, 08, 16, 14, 04 \\
2 & Task 2: 08, 07, 12, 05, 14, 04, 16 & Task 1: 16, 07, 05, 08, 14, 12, 04 & Task 3: 12, 14, 05, 16, 04, 07, 08 \\
3 & Task 2: 12, 16, 04, 05, 07, 14, 08 & Task 1: 04, 12, 07, 14, 16, 08, 05 & Task 3: 08, 12, 14, 07, 05, 16, 04 \\
4 & Task 3: 05, 08, 04, 12, 07, 16, 14 & Task 2: 07, 04, 14, 05, 08, 16, 12 & Task 1: 16, 08, 12, 05, 07, 04, 14 \\
5 & Task 2: 04, 05, 16, 14, 07, 12, 08 & Task 3: 16, 07, 05, 08, 12, 14, 04 & Task 1: 08, 04, 05, 12, 14, 07, 16 \\
6 & Task 2: 12, 04, 05, 07, 16, 14, 08 & Task 3: 05, 08, 14, 16, 07, 12, 04 & Task 1: 14, 05, 04, 12, 07, 16, 08 \\
7 & Task 3: 07, 05, 14, 08, 04, 16, 12 & Task 1: 07, 04, 12, 08, 14, 05, 16 & Task 2: 05, 08, 14, 04, 12, 16, 07 \\
8 & Task 3: 12, 05, 07, 14, 16, 08, 04 & Task 2: 08, 14, 04, 16, 07, 05, 12 & Task 1: 05, 07, 04, 16, 14, 12, 08 \\
9 & Task 1: 04, 08, 16, 05, 12, 07, 14 & Task 2: 14, 16, 12, 07, 05, 04, 08 & Task 3: 14, 12, 07, 16, 05, 04, 08 \\
10 & Task 1: 14, 07, 16, 05, 04, 08, 12 & Task 2: 07, 14, 12, 04, 16, 08, 05 & Task 3: 16, 12, 07, 08, 05, 14, 04 \\
11 & Task 3: 07, 12, 08, 04, 14, 05, 16 & Task 2: 12, 07, 16, 04, 08, 14, 05 & Task 1: 14, 05, 04, 16, 07, 12, 08 \\
12 & Task 2: 12, 08, 05, 16, 07, 14, 04 & Task 3: 07, 14, 05, 16, 08, 04, 12 & Task 1: 05, 14, 12, 08, 16, 07, 04 \\
13 & Task 2: 07, 04, 08, 16, 12, 14, 05 & Task 3: 14, 08, 12, 05, 07, 16, 04 & Task 1: 08, 05, 12, 04, 14, 07, 16 \\
14 & Task 1: 04, 07, 14, 16, 12, 08, 05 & Task 2: 12, 07, 14, 16, 04, 05, 08 & Task 3: 12, 16, 08, 07, 04, 14, 05 \\
\bottomrule
\end{tabular}
\end{table*}

\subsection{Per-condition estimation of the spatial filters}
\label{sec:app_within_task}

\par
\begin{figure}[htbp]
    \centering
    \includegraphics[width=0.65\linewidth]{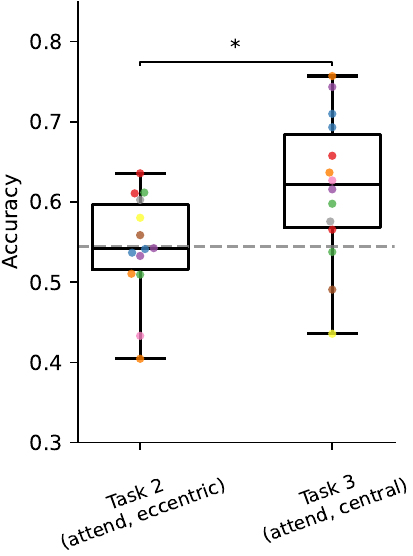}
    \caption{Match-mismatch (MM) decoding accuracy for the attend-eccentric (Task 2) and attend-central (Task 3) conditions on 45-second test segments, obtained with CCA spatial filters estimated separately within each task. This differs from Figure~\ref{fig:acc_tasks}, where the filters are estimated from the combined data of all three tasks. Individual participant accuracies are shown as colored dots, with statistical outliers circled (identified by the Tukey $1.5\times$IQR rule; outliers were retained in all Wilcoxon signed-rank tests). Box plots display the median and interquartile range, with whiskers extending to extreme values. The dashed line indicates the significance bound for above-chance performance, defined as the 97.5th percentile of the circular permutation null distribution recomputed for the present per-condition decoders (1400 permutations). The asterisk denotes a significant difference between tasks (Wilcoxon signed-rank test, $p = 0.028$, $Z=1.91$, rank-biserial $r=0.58$). Task 3 accuracy exceeds Task 2 accuracy, confirming that the eccentricity effect persists under within-task filter estimation.}
    \label{fig:acc_tasks_supp}
\end{figure}

 In the main analysis (Section~\ref{sec:tasks_comparison}), a single set of CCA filters was estimated from the combined recordings of all three tasks and then reused to decode the held-out segments of each individual task (Section~\ref{sec:params_eval}). Since every condition is decoded through the same filters, the comparisons between tasks favor none of them. One may nonetheless ask whether the ranking of the tasks is sensitive to how the filters are obtained: a filter learned from a blend of central and eccentric recordings is a compromise solution that may fit the central and the eccentric conditions to differing degrees. To examine this, we re-estimated the filters separately for Tasks 2 and 3. For each task, a dedicated filter was learned and tested only on the recordings of that same task, again with leave-one-video-out cross-validation, so that the decoder for each condition was derived exclusively from that condition's own data. This variant draws on a third of the training material used in the main analysis, so the resulting accuracies are inevitably lower (Figure \ref{fig:acc_tasks_supp}). What matters here is therefore the relative ordering of Tasks 2 and 3 rather than the absolute accuracy levels.

\par Even under this stricter, condition-specific estimation, decoding accuracy for the central condition (Task 3) remained above that for the eccentric condition (Task 2) (Wilcoxon signed-rank test, $p = 0.028$, $Z=1.91$, rank-biserial $r=0.58$). The eccentricity effect from Section~\ref{sec:tasks_comparison} thus holds under an entirely independent estimation scheme, showing that the weaker neural tracking of motion at peripheral locations originates in the neural responses themselves and is not an incidental by-product of how the spatial filters were learned.

\bibliographystyle{IEEEbib}
\bibliography{refs}

\end{document}